\documentclass[preprint,prd,aps,nofootinbib]{revtex4}
\usepackage{graphicx}
\usepackage{float}
\usepackage{diagbox}
\usepackage{color}
\usepackage{slashed}
\makeatletter

\newcommand{\Rmnum}[1]{\expandafter\@slowromancap\romannumeral #1@}
\makeatother
\begin{document}
\title{The decay property of the $X(3842)$ as the $\psi_{_3}(1^3D_{_3})$ state}
\author{Wei Li$^{1,2,3}$\footnote{watliwei@163.com, corresponding author},
 Su-Yan Pei$^{1,2,3}$, Tianhong Wang$^{4}$,
 Tai-Fu Feng$^{1,2,3}$,
 Guo-Li Wang$^{1,2,3}$\footnote{wgl@hbu.edu.cn, corresponding author}}

\affiliation{&&${^1}$ Department of Physics, Hebei University, Baoding 071002, China
\nonumber\\$^{2}$ Hebei Key Laboratory of High-precision Computation and Application of Quantum Field Theory, Baoding 071002, China
\nonumber\\$^3$ Hebei Research Center of the Basic Discipline for Computational Physics, Baoding 071002, China
\nonumber\\$^{4}$School of Physics, Harbin Institute of Technology, Harbin 150001, China
}
\begin{abstract}
In this paper, the new particle $X(3842)$ discovered by the LHCb Collaboration is identified to be the $\psi_{_3}(1^3D_{_3})$ state. We study its strong decays with the combination of the Bethe-Salpeter method and the $^3P_{_0}$ model. Its electromagnetic (EM) decay is also calculated by the Bethe-Salpeter method within Mandelstam formalism. The strong decay widths are {$\Gamma[X(3842)\rightarrow D^{0}\bar{D}^{0}]=1.28$ MeV}, $\Gamma[X(3823)\rightarrow D^{+}D^{-}]=1.08$ MeV, and the ratio ${\cal B}[X(3842)\rightarrow D^{+}D^{-}]/{\cal B}[X(3823)\rightarrow D^{0}\bar{D}^{0}]=0.84$. The EM decay width is $\Gamma[X(3842)\rightarrow\chi_{_{c2}}\gamma]=0.29$ MeV. We also estimate the total width to be 2.87 MeV, which is in good agreement with the experimental data $2.79^{+0.86}_{-0.86}$ MeV. Since the used relativistic wave functions include different partial waves, we also study the contributions of different partial waves in electromagnetic decay.
\end{abstract}
\maketitle
\section{INTRODUCTION}

It is a known fact that spectra of charm mesons have been experimentally mapped with great precision since the discovery of the $J/\psi$ resonance \cite{E598collaboration1974,SLAC1974}. Theoretically, the potential models \cite{E.Eichten1978} can well describe the spectra and properties of these states. Charmonium, a bound state composed of charm and anti-charm quarks, which is useful for testing the validity of phenomenological models, such as the quark potential model \cite{S.G1985}, which have foreseen rich and meaningful quarkonium spectra, plays an important role in quantum chromodynamics (QCD). In recent decade, the Belle, BABAR and BESIII Collaborations have observed many new Charmonium-like states, commonly known as the $XYZ$ states \cite{K.A.Olive2022}, such as the $X(3872)$ \cite{S.K.Choi2003}, $X(3930)$ \cite{Bell2006,BABAR2010}, $X(3940)$ \cite{Bell2007,Bell2008}, $X(3915)$ \cite{BABAR2012}, $X(3860)$ \cite{Belle2017}, etc. Some of them are traditional excited charmonia, others are considered to be exotic in nature. Interest in charmonium spectroscopy was renewed as more these states were discovered.

Recently, the LHCb Collaboration discovered a new narrow but very high statistical significance resonance state, named $X(3842)$, in the decay modes $X(3842)\rightarrow D^{0}\bar{D}^{0}$ and $X(3842)\rightarrow D^{+}D^{-}$ \cite{R.Aaij2019}. The mass and width of this state are measured to be
$$M_{_{X(3842)}}=3842.71\pm0.16\pm0.12~\rm{MeV},~~~~~~\Gamma_{_{X(3842)}}=2.79\pm0.51\pm0.35~\rm{MeV},$$
where the first uncertainty is statistical and the second is systematic. Based on observed mass and narrow natural width, this new state can be interpreted as the unobserved $\psi_{_3}(1^3D_{_3})$ charmonium state with $J^{PC}=3^{--}$. The BESIII Collaboration confirmed this particle in the process $e^{+}e^{-}\rightarrow \pi^{+}\pi^{-}X(3842)\rightarrow \pi^{+}\pi^{-}D^{+}D^{-}$, and evidence with a significance of $4.2\sigma$ is found  \cite{M.Ablikim2022}.

At present, the experimental data of X(3842) is relatively sparse. However, theory had predicted the $\psi_{_3}(1^3D_{_3})$ state to have a natural width $0.5\sim4$ MeV \cite{E.J.Eichten2006,T.Barnes2004,T.Barnes2005}, and the mass in the rang $3806\sim3912$ MeV \cite{E.J.Eichten1981,S.G1985,S.N.Gupta1986,L.P.Fulcher1991,J.Zeng1995,D.Ebert2003,G.L.Yu2019,S.P2019}. These studies shows that its dominate decay channel is decay to $D^{+}D^{-}$ and $D^{0}\bar{D}^{0}$. In addition, the radiative decay of $X(3842)$ decay to $\chi_{_{c2}}\gamma$ is not negligible \cite{T.Barnes2005}. As is well known, these open-flavor strong decays closely relate to the non-perturbative properties, and our knowledge is rather poor in this region. A complete understanding of the QCD vacuum is necessary to fully solve this problem. Although we can expect lattice QCD calculation to provide us with more reliable theoretical predictions in the future, but for now, we still need to build phenomenological models to study properties of this kind decay, e.g. the $^3P_{_0}$ model \cite{L.Micu1969,A.Le1973,A.Le1974}, the flux-tube mode \cite{R.Kokoski1987}, cornell model \cite{E.Eichten1978,E.E1980} with a vector confinement interaction, the model in Ref.\cite{E.S.A1996} with a scalar confinement interaction, field correlator method \cite{Yu.A2008}.

In previous studies \cite{geng,wangwu}, we found that the relativistic effect of a highly excited state is very large, we need to choose the relativistic method to calculate.
The Salpeter equation \cite{E.E.S1952} is instantaneous version of the Bathe-Salpeter (BS) equation \cite{E.E.S AND H.A.B.1951}, it is suitable for the heavy meson, especially the double-heavy meson. We have solved the complete Salpeter equations for different states, see Refs. \cite{C.S.Kim2004,G.LWang2006,G.L.Wang2007,G.LWang2009}, or the summary papers \cite{C.Hsi.Chang2010,G.L.Wang2022}. We have also improved this method to calculate the transition amplitude \cite{C.Hsi.Chang2006} with relativistic wave function as input. Using this improved BS method, we can get relatively accurate theoretical results, which are in good agreement with the experimental data, see Refs. \cite{fhfeng2011,liqiang2017,Z.H.Wang2022} for examples. The $^3P_{_0}$ model (Quark Pair Creation Model, QPC) is a non-relativistic model. This model is widely used in the Okubo-Zweig-Iizuka (OZI) allowed strong decays of a meson \cite{T.Barnes2005,E.S.A1996,F.E.Close2005}. In Refs. \cite{H.F.Fu2012,T.h.Wang2013,S.C.Li2018}, the $^3P_{_0}$ model is extended to the relativistic case, where the input relativistic wave functions come from the strict solution of the Salpeter equation. So in this paper, the strong decay of $X(3842)$ as $\psi_{_3}(1^3D_{_3})$ state is studied by the combination of the Salpeter equation and the $^3P_{_0}$ model, and its main EM decay is also studied by the improved BS method. In addition, since the relativistic wave function contains different partial waves \cite{G.L.Wang2022}, we also study the contributions of different partial waves in EM decay.

The paper is organized as following. In Sec \Rmnum{2}, We show the relativistic wave functions of initial and final mesons. The formula to calculate the strong and EM decay of $X(3842)$ are also present in this section. In Sec \Rmnum{3}, we give the results and make comparisons with other theoretical predictions and experimental data. Finally, we give the discussion and conclusion.

\section{ THE THEORETICAL CALCULATIONS }
For the sake of brevity of paper, the detailed introduction to the BS equation and $^3P_{_0}$ model, as well as their combination is not provided here. Interested readers may refer to \cite{E.E.S1952,E.E.S AND H.A.B.1951,L.Micu1969,A.Le1973,A.Le1974} or our previous papers, for example, \cite{H.F.Fu2012,T.h.Wang2013}.

\subsection{Transition amplitude of strong decay}

The $^3P_{_0}$ model \cite{L.Micu1969} is non-relativistic. Its core idea is that the quark and anti-quark pair $q\bar{q}$ pairs excited by the operator $g\int d\vec{x}\bar{\psi}\psi\mid_{_{nonrel}}$ \cite{E.S.Swanson2006} from vacuum, carry vacuum quantum number namely $J^{PC}=0^{++}$, which corresponds to a pair of quark and anti-quark with $^{2L+1}L_{_J}= {^3P_{_0}}$ quantum number, so it is called $^3P_{_0}$ model. In order to combine with the BS wave function, we extend this operator to the relativistic covariation form $-ig\int d^4x\bar{\psi}\psi$ (a similar form of interaction is also used in Refs. \cite{Yu.A2008,I.V.Danilkin2010}). Where $g$ can be written as $2m_{_q}\gamma$ and $m_{_q}$ is the constitute quark ($u, d, s$) mass. $\gamma$ is the dimensionless interaction strength, we take $\gamma=0.4$ in this paper.

\begin{figure}[!htb]
\centering
\includegraphics[width=3.9in]{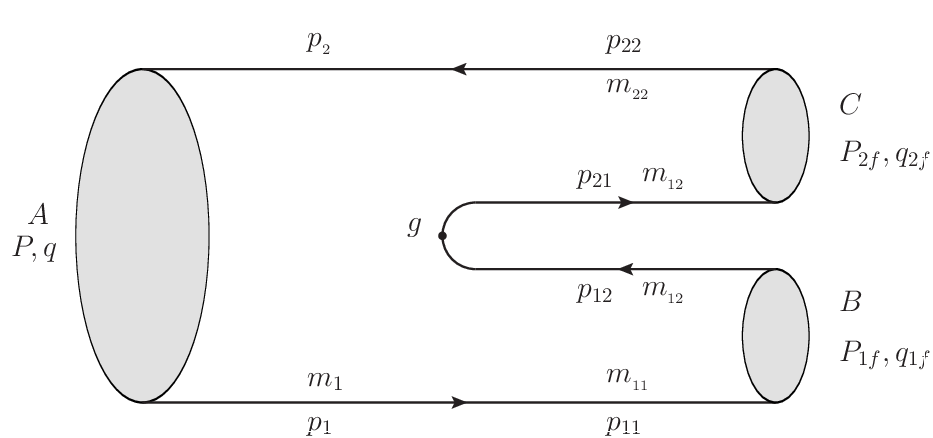}
\caption{The Feynman diagram for the two-body strong decay process with a $^3P_{_0}$ vertex.}
\label{SDFeymP}
\end{figure}

For the two-body strong decay $A\rightarrow B+C$, the Feynman diagram is shown in Fig. \ref{SDFeymP}.
Within the Mandelstam formalism \cite{Mandelstam}, the transition amplitude of strong decay $A\rightarrow B+C$ process can be written as
\begin{eqnarray}\label{am1}
&&\langle P_{_{1f}}P_{_{2f}}\mid S \mid P\rangle_{_{^3P_{_0}}}=(2\pi)^4\delta^4(P-P_{_{1f}}-P_{_{2f}}){\cal M}_{_{^3P_{_0}}}
\nonumber
\\&&=-ig(2\pi)^4\delta^4(P-P_{_{1f}}-P_{_{2f}})
\int\frac{d^4q}{(2\pi)^4}Tr\bigg[\chi_{_P}(q)S^{-1}_{_2}(-p_{_{2}})
\bar{\chi}_{_{P_{_{2f}}}}(q_{_{2f}})\bar{\chi}_{_{P_{_{1f}}}}(q_{_{1f}})S^{-1}_{_1}(p_{_{1}})\bigg],
\end{eqnarray}
where $\chi_{_P}(q)$, ${\chi}_{_{P_{1f}}}(q_{_{1f}})$, ${\chi}_{_{P_{2f}}}(q_{_{2f}})$ are the relativistic BS wave functions for initial and final mesons, respectively. The internal relative momenta of the initial and final mesons are $q$, $q_{_{1f}}$ and $q_{_{2f}}$, respectively. $p_{_1}$, $p_{_2}$, $S_{_1}(p_{_{1}})$ and $S_{_2}(-p_{_{2}})$ represent the momenta and propagators of quark and anti-quark, respectively.

Since we solve the complete Salpeter equation, not the BS equation, so the instantaneous approximation has been used to BS equation, and the Salpeter wave functions are obtained. So we make instantaneous approximation to the upper amplitude, namely, integrate over the $q_0$. Then we obtain the transition amplitude with the Salpeter wave functions as input \cite{H.F.Fu2012,T.h.Wang2013},

\begin{eqnarray}\label{SD}
{\cal M}_{_{^3P_{_0}}}=g\int\frac{d^3q_{_\perp}}{(2\pi)^3}Tr\bigg[\frac{\slashed{P}}{M}\varphi^{++}_{_P}
(q_{_\perp})
\frac{\slashed{P}}{M}\bar{\varphi}^{++}_{_{P_{_{2f}}}}(q_{_{2f_\perp}})\bar{\varphi}^{++}_{_{P_{_{1f}}}}
(q_{_{1f_\perp}})\bigg]
\bigg(1-\frac{M-\omega_{_{1}}-\omega_{_{2}}}{2\omega_{_{12}}}\bigg),
\end{eqnarray}
where $M$ is the mass of initial state, $\varphi^{++}_{_P}$ and $\varphi^{++}_{_{P_{_{if}}}}$ is the positive energy wave function of initial and final mesons, respectively. We have defined $q^{\mu}_{_{\perp}}=q^{\mu}-(P\cdot{q}/M^{2})P^{\mu}$. The relation between the relative momentua in  initial and final mesons are $q_{_{if}}=q+(-1)^{i+1}(\alpha_{_{i}}P-\alpha_{_{ii}}P_{_{if}})$, where $i=1,2$, $\alpha_{_{i}}=\frac{m_{_i}}{m_{_1}+m_{_2}}$, $\alpha_{_{ii}}=\frac{m_{_{ii}}}{m_{_{i1}}+m_{_{i2}}}$, subscript $f$ means this quantity belongs to the final state. $\omega_{_{i}}=\sqrt{m^2_{_{i}}-q^2_{_{{\perp}}}}$ and $\omega_{_{12}}\equiv\sqrt{m^2_{_{12}}-p^2_{_{12\perp}}}$.
Considering $M\simeq\omega_{_{1}}+\omega_{_{2}}$, the second term in parentheses can be ignored (when $|q_{_{\perp}}|$ is large, this approximation is not true, but at this time, the value of wave function is also very small, thus greatly reducing the contribution of this part).

\subsection{Transition amplitude of EM decay}

\begin{figure}[!htb]
\begin{minipage}[c]{1\textwidth}
\includegraphics[width=3in]{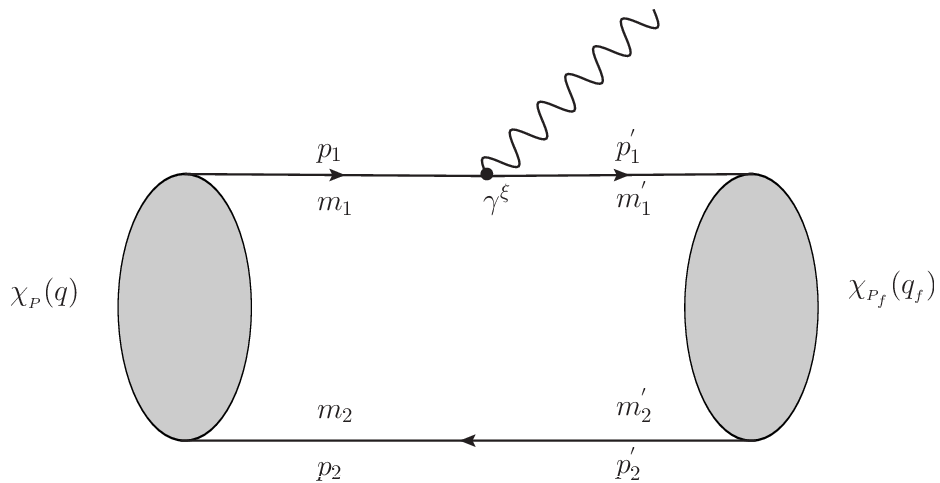}
\includegraphics[width=3in]{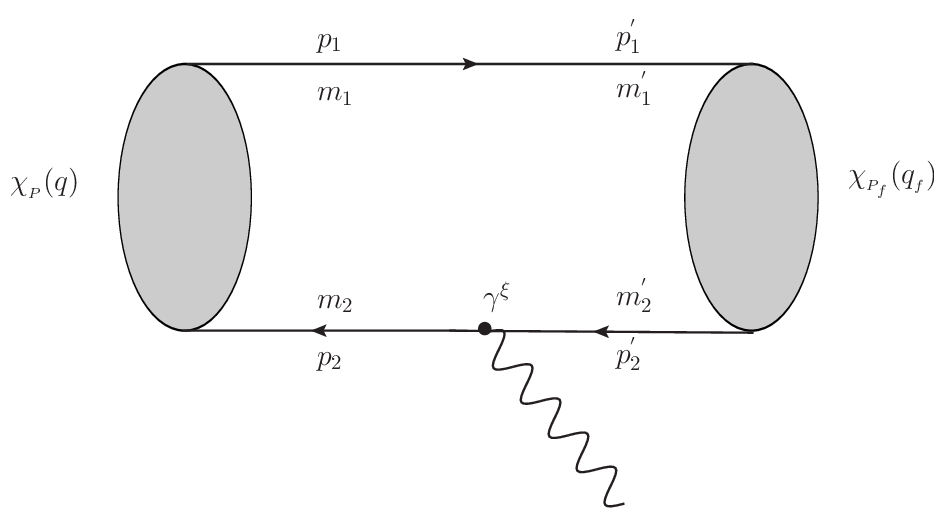}
\end{minipage}
\caption{Feynman diagram for the transition $\chi_{_P}\rightarrow\chi_{_{cJ}}\gamma$. The diagram on the left and right show photons come from the quark and the anti-quark, respectively.}
\label{feynman}
\end{figure}
For the EM decay of $A\rightarrow B\gamma$, the transition amplitude can be written as
\begin{eqnarray}
&&\langle \chi_{_{cJ}}(P_{_f},\epsilon_{_2})\gamma(k,\epsilon_{_0})|X(P,\epsilon_{_1})\rangle=(2\pi)^{4}\delta^{4}
(P-P_{_f}-k)\epsilon_{_{0{\xi}}}{\cal M}^{\xi},
\end{eqnarray}
where $\epsilon_{_0}$, $\epsilon_{_1}$ and $\epsilon_{_2}$ are the polarization vectors (tensor) of the photon, initial and final mesons, respectively. $P$, $P_{_f}$ and $k$ are the momenta of initial meson, final meson and photon, respectively. {
From the quantum number perspective, we know that the electromagnetic processes of $^3D_3\to{}^3P_J\gamma$ ($J=0,1,2$) are all $E_1$ dominant decays. But in non-relativistic limit \cite{W.Kwong1988,T.Barnes2004}, the $E_1$ decay widths of $X(3842)$ to $\chi_{_{c0}}\gamma$ and $\chi_{_{c1}}\gamma$ are zero, so they are actually the $M_2$ dominant decays \cite{W.Li2023}, and have very small partial widths. In addition, the channels $X(3842)\to\eta_{_{c}}\gamma$ and $X(3842)\to c\bar c(^1D_2)\gamma$ are $M_1$ dominant modes, and will not have large partial widths.
Therefore, in this paper, we only calculate the $E_1$ dominant channel $X(3842)\rightarrow \chi_{_{c2}}\gamma$, and ignore other electromagnetic modes.}

From the Fig. \ref{feynman}, we can see that invariant amplitude ${\cal M}^{\xi}$ consists of two parts, where photons are emitted from quark and anti-quark, respectively. In the condition of instantaneous approach, the amplitude can be written as \cite{C.Hsi.Chang2006}
\begin{eqnarray}\label{em}
&&{\cal M}^{\xi}=\int{\frac{d^3q_{_\perp}}{(2\pi)^{3}}}Tr\bigg[Q_{_1}e\frac{\slashed{P}}{M}\bar{\varphi}^{++}_{_f}
(q_{_\perp}+\alpha_{_2}P_{_{f_\perp}})
\gamma^{\xi}\varphi^{++}_i(q_{_\perp})
\nonumber\\&&\hspace{1.3cm}+Q_{_2}e~\bar{\varphi}^{++}_{_f}(q_{_\perp}-\alpha_{_1}P_{_{f_\perp}})
\frac{\slashed{P}}{M}\varphi^{++}_{_i}(q_{_\perp})
\gamma^{\xi}\bigg],
\end{eqnarray}
where $Q_{_1}$ and $Q_{_2}$ are the electric charges (in unit of $e$) of quark and anti-quark, respectively. $\varphi^{++}_{_{i,f}}$ is the positive energy wave function, $i,f$ stand for initial and final states, respectively.

\subsection{The relativistic wave functions}

In the calculation, we use the relativistic Salpeter wave function for $X(3842)$, which is a $3^{--}$ state \cite{Th.Wang2016},
\begin{eqnarray}\label{3--}
&&\varphi_{_{3^{--}}}(q_{_\perp})=\epsilon_{_{\mu\nu\alpha}}q^{\mu}_{_\perp}q^{\nu}_{_\perp}
\bigg[q^{\alpha}_{_\perp}\bigg(f_{_1}+\frac{\slashed{P}}{M}f_{_2}
+\frac{\slashed{q}_{\perp}}{M}f_{_3}+\frac{\slashed{P}\slashed{q}_{\perp}}{M^2}f_{_4}\bigg)
\nonumber\\&&\hspace{2cm}+M\gamma^{\alpha}\bigg(f_{_5}+\frac{\slashed{P}}{M}f_{_6}
+\frac{\slashed{q}_{\perp}}{M}f_{_7}+\frac{\slashed{P}\slashed{q}_{\perp}}{M^2}f_{_8}\bigg)\bigg],
\end{eqnarray}
where, $\epsilon_{_{\mu\nu\alpha}}$ is the third-order polarization tensor of the state $X(3842)$. Radial wave function $f_{_i}$ $(i=1,2,...8)$ is function of $-q^2_{_\perp}$, and its numerical value will be obtained by solving the Salpeter equation. In our method, not all radial wave functions $f_{_i}s$ are independent, only four of them are. The relationships between them are given in the Appendix A.

{We now show that every terms in $3^{--}$ state wave function have negative parity and negative charge conjugate parity.
When we perform parity transformation, $P'=(P_0,-\vec{P})$, $q'=(q_0,-\vec{q})$, and set
 $$\varphi_{_{3^{--}}}(P,q_{_\perp})=\eta_{_P}\gamma_{_0}\varphi{'}_{_{3^{--}}}(P',q{'}_{_\perp})\gamma_{_0},$$
where $\eta_{_P}$ is the parity.
In the center of mass system, we have $P'=P$, $q{'}_{_\perp}=-q_{_\perp}$, and
\begin{eqnarray}\label{PP}
&&\gamma_{_0}\varphi{'}_{_{3^{--}}}(q{'}_{_\perp})\gamma_{_0}=
\epsilon_{_{\mu\nu\alpha}}q^{\mu}_{_\perp}q^{\nu}_{_\perp}
\bigg[-q^{\alpha}_{_\perp}\gamma_{_0}\bigg(f_{_1}+\frac{\slashed{P}}{M}f_{_2}
-\frac{\slashed{q}_{\perp}}{M}f_{_3}-\frac{\slashed{P}\slashed{q}_{\perp}}{M^2}f_{_4}\bigg)\gamma_{_0}
\nonumber\\&&\hspace{4.6cm}+M\gamma_{_0}\gamma^{\alpha}\bigg(f_{_5}+\frac{\slashed{P}}{M}f_{_6}
-\frac{\slashed{q}_{\perp}}{M}f_{_7}-\frac{\slashed{P}\slashed{q}_{\perp}}{M^2}f_{_8}\bigg)\gamma_{_0}\bigg]
\nonumber\\&&\hspace{2.5cm}=-\varphi_{_{3^{--}}}(P,q_{_\perp}),
\end{eqnarray}
so $\eta_{_P}=-1$.

When we take charge conjugate transformation,  $$\varphi_{_{3^{--}}}(q_{_\perp})=\eta_{_C}C\varphi^T_{_{3^{--}}}(-q_{_\perp})C^{-1},$$
where $T$ is the rotation transform, $C$ is the charge conjugate transform, $C{\gamma_{\mu}^T}C^{-1}=-\gamma_{\mu}$, $\eta_{_C}$ is the charge conjugate parity. We have
\begin{eqnarray}\label{CP}
&&C\varphi^{T}_{_{3^{--}}}(-q_{_\perp})C^{-1}=C\bigg\{\epsilon_{_{\mu\nu\alpha}}q^{\mu}_{_\perp}q^{\nu}_{_\perp}
\bigg[-q^{\alpha}_{_\perp}\bigg(f_{_1}+\frac{\slashed{P}^T}{M}f_{_2}
-\frac{\slashed{q}^T_{\perp}}{M}f_{_3}-\frac{\slashed{q}^T_{\perp}\slashed{P}^T}{M^2}f_{_4}\bigg)
\nonumber\\&&\hspace{5.5cm}+M\bigg(f_{_5}+\frac{\slashed{P}^T}{M}f_{_6}
-\frac{\slashed{q}^T_{\perp}}{M}f_{_7}-\frac{\slashed{q}^T_{\perp}\slashed{P}^T}{M^2}f_{_8}\bigg)\gamma^{\alpha T}\bigg]\bigg\}C^{-1}
\nonumber\\&&\hspace{3.2cm}=-\epsilon_{_{\mu\nu\alpha}}q^{\mu}_{_\perp}q^{\nu}_{_\perp}
\bigg[q^{\alpha}_{_\perp}\bigg((f_{_1}+2f_{_7})+\frac{\slashed{P}}{M}(2f_{_8}-f_{_2})+\frac{\slashed{q}_{\perp}}{M}f_{_3}+\frac{\slashed{P}\slashed{q}_{\perp}}{M^2}f_{_4}\bigg)
\nonumber\\&&\hspace{6.4cm}+M\gamma^{\alpha}\bigg(f_{_5}+\frac{\slashed{P}}{M}f_{_6}
-\frac{\slashed{q}_{\perp}}{M}f_{_7}+\frac{\slashed{P}\slashed{q}_{\perp}}{M^2}f_{_8}\bigg)\bigg]
\nonumber\\&&\hspace{3.2cm}=-\varphi_{_{3^{--}}}(q_{_\perp}),
\end{eqnarray}
so $\eta_{_C}=-1$. Since only quarkonium has $C$ parity, we have used $m_{_1}=m_{_2}$, $\omega_{_1}=\omega_{_2}$, $f_{_2}=f_{_8}$ and $f_{_7}=0$ in Eq.(\ref{CP}), see Appendix A.

Further, we show that the relativistic wave function in Eq.(\ref{3--}) for $3^{--}$ state is not a pure $D$-wave. In terms of spherical harmonics $Y_{_{lm}}$, we can rewrite
\begin{eqnarray}\label{12}
&&\epsilon_{_{\mu\nu\alpha}}q^{\mu}_{_\perp}q^{\nu}_{_\perp}q^{\alpha}_{_\perp}=2i\sqrt{\frac{6\pi}{35}}\mid\vec{q}\mid^3(Y_{_{32}}-Y_{_{3-2}}),
\end{eqnarray}
so $f_{_1}$ and $f_{_2}$ terms are $F$-wave. Similarly, for the $f_{_3}$ and $f_{_4}$ terms,
\begin{eqnarray}\label{34}
&&\epsilon_{_{\mu\nu\alpha}}q^{\mu}_{_\perp}q^{\nu}_{_\perp}q^{\alpha}_{_\perp}\slashed{q}_{\perp}=i\mid\vec{q}\mid^4
\bigg[\frac{4}{7}\sqrt{\frac{3\pi}{5}}(-Y_{_{21}}\gamma^{+}+Y_{_{2-1}}\gamma^{-})+\frac{2}{7}\sqrt{\frac{6\pi}{5}}(-Y_{_{22}}+Y_{_{2-2}})\gamma^{\Delta}
\nonumber\\&&+\frac{2}{7}\sqrt{\frac{2\pi}{5}}(Y_{_{41}}\gamma^{+}-Y_{_{4-1}}\gamma^{-})+\frac{4}{7}\sqrt{\frac{2\pi}{5}}(-Y_{_{42}}+Y_{_{4-2}})\gamma^{\Delta}
+2\sqrt{\frac{2\pi}{35}}(Y_{_{43}}\gamma^{-}-Y_{_{4-3}}\gamma^{+})\bigg].
\end{eqnarray}
where, $\gamma^{+}=-\frac{\gamma^{1}+i\gamma^{2}}{\sqrt{2}}$, $\gamma^{-}=\frac{\gamma^{1}-i\gamma^{2}}{\sqrt{2}}$ and $\gamma^{\Delta}=\gamma^{3}$. So $f_{_3}$ and $f_{_4}$ terms include $D$-wave and $G$-wave, they are $D-G$ mixing. The pure $G$-wave in Eq.(\ref{3--}) is
$$\epsilon_{_{\mu\nu\alpha}}q^{\mu}_{_\perp}q^{\nu}_{_\perp}
q^{\alpha}_{_\perp}(\frac{\slashed{q}_{\perp}}{M}f_{_3}+\frac{\slashed{P}\slashed{q}_{\perp}}{M^2}f_{_4})
+\frac{3}{7}\epsilon_{_{\mu\nu\alpha}}q^{\mu}_{_\perp}q^{\nu}_{_\perp}M\gamma^{\alpha}\frac{\mid\vec{q}\mid^2}{M^2}(f_{_3}-\frac{\slashed{P}}{M}f_{_4}).$$
The $f_{_5}$ and $f_{_6}$ terms are $D$ wave, because,
\begin{eqnarray}\label{56}
&&\epsilon_{_{\mu\nu\alpha}}q^{\mu}_{_\perp}q^{\nu}_{_\perp}\gamma^{\alpha}_{_\perp}=2i\sqrt{\frac{2\pi}{15}}\mid\vec{q}\mid^2\bigg[\sqrt{2}(Y_{_{21}}\gamma^{+}-Y_{_{2-1}}\gamma^{-})
+(Y_{_{22}}-Y_{_{2-2}})\gamma^{\Delta}\bigg].
\end{eqnarray}
Then the complete $D$-wave in Eq.(\ref{3--}) is
$$\epsilon_{_{\mu\nu\alpha}}q^{\mu}_{_\perp}q^{\nu}_{_\perp}M\gamma^{\alpha}(f_{_5}+\frac{\slashed{P}}{M}f_{_6})
-\frac{3}{7}\epsilon_{_{\mu\nu\alpha}}q^{\mu}_{_\perp}q^{\nu}_{_\perp}M\gamma^{\alpha}\frac{\mid\vec{q}\mid^2}{M^2}(f_{_3}-\frac{\slashed{P}}{M}f_{_4}).$$
The $f_{_7}$ and $f_{_8}$ terms are $F$ wave, since
\begin{eqnarray}\label{78}
&&\epsilon_{_{\mu\nu\alpha}}q^{\mu}_{_\perp}q^{\nu}_{_\perp}\gamma^{\alpha}_{_\perp}\slashed{q}_{\perp}=2i\mid\vec{q}\mid^3
\bigg[\frac{2}{5}\sqrt{\frac{\pi}{7}}Y_{_{30}}({\gamma^{+}}^2-{\gamma^{-}}^2)+\frac{1}{5}\sqrt{\frac{6\pi}{7}}(-Y_{_{31}}\gamma^{+}+Y_{_{3-1}}\gamma^{-})\gamma^{\Delta}
\nonumber\\&&\hspace{2.3cm}+\sqrt{\frac{2\pi}{105}}(Y_{_{32}}-Y_{_{3-2}})(2\gamma^{+}\gamma^{-}
-{\gamma^{\Delta}}^2)+\sqrt{\frac{2\pi}{35}}(Y_{_{33}}\gamma^{-}-Y_{_{3-3}}\gamma^{+})\gamma^{\Delta}\bigg].
\end{eqnarray}

So, it can be concluded that from our relativistic wave function that the widely used representations of $P=(-1)^{L+1}$ and $C=(-1)^{L+S}$, as well as $^{2S+1}L_J$, can only be strictly used in a non-relativistic condition.
}

For strong decay final states, they can only be $0^{-}$ state pseudoscalars $D^0$ and $\bar{D^0}$ (or $D^+$ and $D^-$).
The relativistic wave function of a $0^{-}$ state can be written as \cite{C.S.Kim2004}
\begin{eqnarray}\label{0-}
&&\varphi_{_{0^{-}}}(q_{_{f_\perp}})=M_{_f}\bigg(\frac{\slashed{P_{_f}}}{M_{_f}}g_{_1}+g_{_2}
+\frac{\slashed{q}_{{f_\perp}}}{M_{_f}}g_{_3}+\frac{\slashed{P_{_f}}\slashed{q}_{{f_\perp}}}{M^2_{_f}}
g_{_4}\bigg)\gamma^{5},
\end{eqnarray}
where the subscript $f$ indicates that the quantity is in the final state. The radial wave functions $g_{_3}$ and $g_{_4}$ are not independent, they are related to $g_{_1}$ and $g_{_2}$, we show their relationship in Appendix A. Similarly, the relativistic $0^-$ state wave function is not a pure $S$-wave (the terms including $g_{_1}$ and $g_{_2}$), but also contains a small component of $P$-wave (terms including $g_{_3}$ and $g_{_4}$).

The final state of EM decay is the charmonium $2^{++}$ state $\chi_{_{c2}}$, and its relativistic wave function can be written as \cite{G.LWang2009}
\begin{eqnarray}\label{2++}
&&\varphi_{_{2^{++}}}(q_{_{f_\perp}})=\epsilon_{_{\mu\nu}}q^{\mu}_{_{f_\perp}}\bigg[q^{\nu}_{_{f_\perp}}
\bigg(h_{_1}+\frac{\slashed{P_{_f}}}{M_{_f}}h_{_2}
+\frac{\slashed{q}_{{f_\perp}}}{M_{_f}}h_{_3}+\frac{\slashed{P_{_f}}\slashed{q}_{{f_\perp}}}
{M^2_{_f}}h_{_4}\bigg)
\nonumber\\&&\hspace{2cm}+M_{_f}\gamma^{\nu}\bigg(h_{_5}+\frac{\slashed{P_{_f}}}{M_{_f}}h_{_6}
+\frac{\slashed{q}_{{f_\perp}}}{M_{_f}}h_{_7}\bigg)+\frac{i}{M_{_f}}h_{_8}\epsilon^{\mu\alpha\beta\gamma}
P_{f_{\alpha}}q_{_{f_{\perp\beta}}}
\gamma_{_\gamma}\gamma_{_5}\bigg],
\end{eqnarray}
where $\epsilon_{_{\mu\nu}}$ is the second-order polarization tensor,  $\epsilon_{_{\mu\nu\alpha\beta}}$ is the Levi-Civita simbol. Four radial wave function $h_{_3}$, $h_{_4}$, $h_{_5}$ and $h_{_6}$ are independent, others are related to them, see the Appendix A. The wave function of $2^{++}$ state is $P$-wave dominant, but is also contains other partial waves. In Eq. (\ref{2++}), the terms including $h_{_5}$ and $h_{_6}$ are $P$-waves, those including $h_{_3}$ and $h_{_4}$ are $P-F$ mixing waves, others are $D$ waves.

\begin{figure}[!htb]
\begin{minipage}[c]{1\textwidth}
\includegraphics[width=2in]{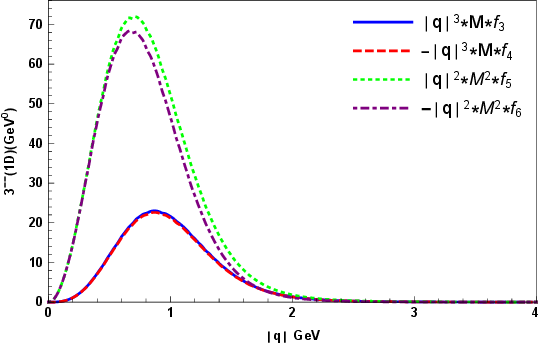}
\includegraphics[width=2in]{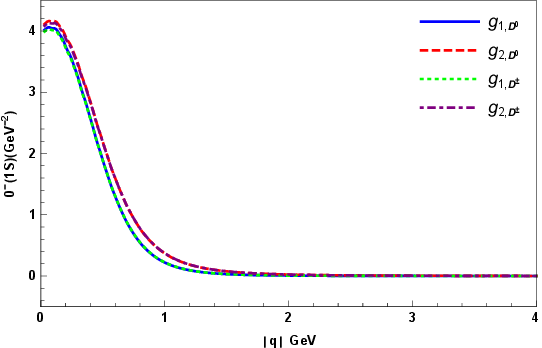}
\includegraphics[width=2in]{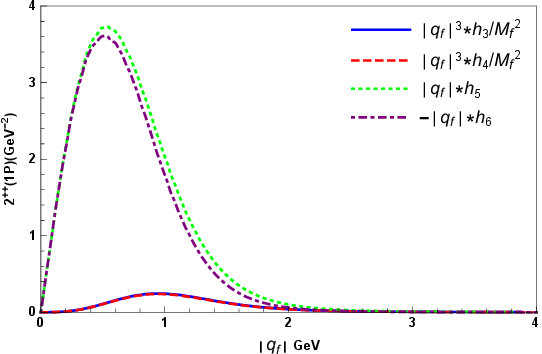}
\end{minipage}
\caption{The radial wave functions of the $3^{--}$ state $X(3842)$, $0^-$ states $D^{+}$ ($D^{-}$) and $D^{0}$ ($\bar{D}^{0}$) , and $2^{++}$ state $\chi_{c2}$, from left to right, respectively.}
\label{rwf}
\end{figure}

We do not show the details of  solving the complete Salpeter equations, only give the numerical values of independent radial wave functions in Fig.\ref{rwf}.
From Fig.\ref{rwf}, we can see that $f_{_5}\simeq -f_{_6}$,
$g_{_1}\simeq g_{_2}$, and $h_{_5}\simeq -h_{_6}$, these equality relations under non-relativistic condition confirm the correctness of our method.

\subsection{The form factors}

Inserting Eq. (\ref{3--}) and Eq. (\ref{0-}) into Eq. (\ref{SD}), where we integrate internal momentum ${q}_{_\perp}$ over the initial and final state wave functions, and finishing the trace, then we obtain  the strong decay amplitude described using form factor,
\begin{eqnarray}\label{sta}
&&{\cal M}(X(3842)\rightarrow D\bar{D})=\epsilon_{_{\mu\nu\alpha}}P^{\mu}_{_f}P^{\nu}_{_f}P^{\alpha}_{_f}t_{_1},
\end{eqnarray}
where $t_{_1}$ is the form factor.

In the same way, inserting Eq. (\ref{3--}) and Eq. (\ref{2++}) into Eq. (\ref{em}), the EM transition amplitude described by form factors are obtained as,
\begin{eqnarray}\label{ema}
&&{\cal M}^{\xi}(X(3842)\rightarrow \chi_{_{c2}}(1P)\gamma)=P^{\xi}\epsilon_{_{P_{f}P_{f}P_{f}}}\epsilon_{_{f,PP}}s_{_1}+
+P^{\xi}\epsilon_{_{\rho~P_{f}P_{f}}}\epsilon^{\rho}_{_{f,P}}s_{_2}+\epsilon^{\xi}_{_{P_{f}P_{f}}}
\epsilon_{_{f,PP}}s_{_3}
\nonumber\\&&\hspace{5.2cm}+\epsilon_{_{P_{f}P_{f}P_{f}}}\epsilon^{\xi}_{_{f,P}}s_{_4}
+P^{\xi}\epsilon_{_{\rho\sigma P_{f}}}\epsilon^{\rho\sigma}_{_{f}}s_{_5}
+\epsilon^{\xi}_{_{\rho~P_{f}}}\epsilon^{\rho}_{_{f,P}}s_{_6}
\nonumber\\&&\hspace{5.2cm}+\epsilon_{_{\rho P_{f}P_{f}}}\epsilon^{\xi\rho}_{_{f}}s_{_{7}}+
\epsilon^{\xi}_{_{\rho\sigma}}\epsilon^{\rho\sigma}_{_{f}}s_{_{8}},
\end{eqnarray}
where $s_{_i}$ is the form factor. We have used some abbreviations, for example, $\epsilon_{_{P_{f}P_{f}P_{f}}}\epsilon_{_{f,PP}}=\epsilon_{_{\mu\nu\alpha}}P^\mu_{_f}P^\nu_{_f}P^\alpha_{_f}
\epsilon_{_{f,\beta\gamma}}P^\beta P^\gamma$. Since the expressions of $t_{_1}$ and $s_{_i}$ are complex, the details are not shown here.

For the EM decay, we note that not all the form factors are independent, due to the Ward identity $(P_{_\xi}-P_{_{f,\xi}}){\cal M}^{\xi}=0$, we have the following relations
\begin{eqnarray}
s_{_3}=(M^2-ME_{_f})s_{_1}+s_{_4},~~ s_{_6}=(M^2-ME_{_f})s_{_2}+s_{_7},~~s_{_8}=(M^2-ME_{_f})s_{_5}.
\end{eqnarray}
The two-body decay width formulation is
\begin{eqnarray}\label{dw}
\Gamma(X\rightarrow{AB})=\frac{|\vec{P_{_f}}|}{8{\pi}M^2}\frac{1}{2J+1}\sum_{\lambda}\overline{{|\cal{M}|}^2},
\end{eqnarray}
where, $|\vec{P_{_f}}|=\sqrt{\left[M^2-(M_{_{1f}}-M_{_{2f}})^2\right]\left[M^2-(M_{_{1f}}+M_{_{2f}})^2 \right]}/2M$ is the three-dimensional momentum of the final meson, $J=3$ is the total angular momentum of initial meson, $\lambda$ represents the polarization of both initial and final mesons.

\section{RESULTS AND DISCUSSIONS}

In our model, the following constituent quark masses are used, $m_{_u}=0.305~\rm{GeV},~m_{_d}=0.311~\rm{GeV},~m_{_c}=1.62~\rm{GeV}$. Other model dependent parameters can be found in Ref. \cite{C.Hsi.Chang2010}, {where we choose the Cornell potential \cite{E.Eichten1978}, a linear scalar potential plus a Coulomb vector potential, since the predicted mass spectrum may not match very well with the experiment data, a free constant parameter $V_{0}$ is usually added to linear scalar potential to fit data \cite{S.G1985}. So by varying the $V_0$ \cite{C.Hsi.Chang2010}, we fit the experimental meson masses $M_{_{X(3842)}}=3.8427~\rm{GeV},~M_{_{D^{+}}}=1.8697~\rm{GeV},~M_{_{D^0}}=1.8648~\rm{GeV},
~M_{_{\chi_{c2}}}=3.5562~\rm{GeV}$ \cite{K.A.Olive2022}, and obtain the numerical values of the corresponding wave functions.}

\subsection{Strong decay widths of $X(3842)$ as $1^3D_{_3}$ state}

The strong decay width of $X(3842)$ decays to $D^{0}\bar{D}^{0}$ and $D^{+}D^{-}$ are calculated as
\begin{eqnarray}
\Gamma[X(3842)\rightarrow D^{0}\bar{D}^{0}]=1.27~\textrm{MeV},~~~ \Gamma[X(3842)\rightarrow D^{+}D^{-}]=1.08 ~\textrm{MeV}.
\end{eqnarray}
For comparison, we show our results and other model predictions \cite{T.Barnes2004,T.Barnes2005,E.J.Eichten2006,G.L.Yu2019,E.J.Eichten2004} and experimental data \cite{R.Aaij2019} in Table \ref{I}. Here we also present the choice of the mass of the initial state $X(3842)$ by the different working groups. Eichten, $et~al$, calculated the decay widths of $1^3D_{_3}\rightarrow DD$ though Refine Conell coupled-channel model \cite{E.J.Eichten2006} and Conell coupled-channel model \cite{E.J.Eichten2004}, respectively. Barnes, $et~al$, used the $^3P_{_0}$ model \cite{T.Barnes2004,T.Barnes2005} to estimated the strong decay width of $1^3D_{_3}(3^{--})$.  Ref. \cite{T.Barnes2004} took the mass $3849$ MeV of $1^3D_{_3}$, which is obtained through the relativistic Godfrey-Isgur model (GI model). The mass $3872$ MeV is also utilized in Ref. \cite{T.Barnes2004} because they consider $X(3872)$ as a candidate for $1^3D_{_3}$. Ref. \cite{T.Barnes2005} used mass $3806$ MeV, which is obtained by Nonrelativistic method (NR-model). Yu $et~al$. \cite{G.L.Yu2019} studied $1^3D_{_3}$ strong decay with QCD sum rules and $^3P_{_0}$ model.
\begin{table}[!htb]
\caption{ The strong decay widths (MeV) of the $X(3842)\rightarrow DD$, and the ratio of $\frac{{\cal B}[X(3842)\rightarrow D^{+}D^{-}]}{{\cal B}[X(3842)\rightarrow D^{0}\bar{D}^{0}]}$.}
\begin{tabular}{c|ccccccc}
\hline
& ~~{\cite{E.J.Eichten2006}}~~ &  ~~ {\cite{T.Barnes2004}}~~ & ~~{\cite{T.Barnes2005}} ~~ & ~~ {\cite{G.L.Yu2019}}~~& ~~{\cite{E.J.Eichten2004}}~~ & ~~ours ~~ &~~EX\cite{R.Aaij2019} ~~\\
\hline
$M_{(1^3D_{_3})}(MeV)$ & $3868$ & $3849$~ $3872$ & $3806$  & $3762\sim3912$  &~$3872$ ~$3902$ & $3842.7$ & $3842.71^{+0.28}_{-0.28}$\\
\hline
$\Gamma(X(3842)\rightarrow D^{+}D^{-})$   &  &  &  & $~2\sim3$ ~& 0.39~~~0.72 & 1.08 & \\
 \hline
 $\Gamma(X(3842)\rightarrow D^{0}\bar{D}^{0})$  & &  &  & ~$2.5\sim3.5$~& 0.47~~~0.84 & 1.27 &  \\
 \hline
${\Gamma(X(3842)\rightarrow DD)}$  & 0.82 & 2.27~~~4.04 & 0.5 & $~4.5\sim6.5$ ~& 0.86~~~1.56 & 2.35 & $2.79^{+0.86}_{-0.86}$ \\
\hline
$\frac{\Gamma(X(3842)\rightarrow D^{+}D^{-})}
{\Gamma(X(3842)\rightarrow D^{0}\bar{D}^{0})}$  &  &   &  & $0.85\sim0.90$ & 0.83~~~0.86 & 0.84 &  \\
\hline
\end{tabular}
\label{I}
\end{table}

From Table \ref{I}, one can see that our prediction of the strong decay width, $2.35$ MeV, is close to the center value of experiment, $2.79^{+0.86}_{-0.86}$ MeV, and consistent well with the result of Ref. \cite{T.Barnes2004}, $2.27$ MeV. And we all used similar masses for $1^3D_{_3}$. In Table \ref{I}, the ratio between two strong decay channels  $X(3842)\rightarrow D^{+}D^{-}$ and $X(3842)\rightarrow D^{0}\bar{D}^{0}$ is also shown. Our result
\begin{eqnarray}
&&\frac{{\cal B}[X(3842)\rightarrow D^{+}D^{-}]}{{\cal B}[X(3842)\rightarrow D^{0}\bar{D}^{0}]}=0.84
\end{eqnarray}
is consistent well with others shown in Table \ref{I}.
{The reason for the good agreement between different theoretical results is simple, as the ratio between the decay width of $X(3842)\rightarrow D^+D^-$ and those of $X(3842)\rightarrow D^0\bar D^0$ is mainly determined by the phase spaces. In Eq.(\ref{sta}), the decay amplitude of $X(3842)\rightarrow D\bar D$ is written as
${\cal M}=\epsilon_{_{\mu\nu\alpha}}P^{\mu}_{_f}P^{\nu}_{_f}P^{\alpha}_{_f}t_{_1}$,
so
$$\sum_{\lambda}\overline{\mid{\cal M}\mid^2}=\frac{2}{5}|\vec{P_{_f}}|^6t^2_{_1}~,$$
the form factor $t_{_1}$ is overlapping integral of initial and final radial wave functions, and from Fig.\ref{rwf}, it can be seen that there is not much difference in the radial wave functions between $D^+$ and $D^0$ mesons. Then using the Eq.(\ref{dw}) of decay width, we obtain
\begin{eqnarray}
&&\frac{\Gamma(X(3842)\rightarrow D^{+}D^{-})}
{\Gamma(X(3842)\rightarrow D^{0}\bar{D}^{0})}\propto \frac{(|\vec{P_{_f}}|_{_{D^{+}}})^7}{(|\vec{P_{_f}}|_{_{D^0}})^7}= 0.735,
\end{eqnarray}
this estimation is very close to our calculated value $0.84$, indicating that the difference between the decay widths of $X(3842)\rightarrow D^0\bar D^0$ and $X(3842)\rightarrow D^+D^-$ is almost purely from the phase space difference, and the isospin symmetry breaking effect is small.}

{Since the very small mass difference $\left(2M_{D^{+}}-2M_{D^0}\right)$ causes a large difference in partial decay widths, and the $1^3D_{_3}$ state $X(3842)$ just lies above the threshold of two charmed mesons, we draw a conclusion that the mass of $X(3842)$ has a significant impact on the value of its strong decay widths. Therefore, accurate experimental mass measurements are crucial for theoretical study on strong decays.}

\subsection{EM decay width of $X(3842)$ as $1^3D_{_3}$ state}

According to Refs. \cite{T.Barnes2005,W.Kwong1988}, only $\psi_{_3}(1^3D_{_3})\rightarrow\chi_{_{c2}}(1P)\gamma$ is dominant in the EM decays of $\psi_{_3}(1^3D_{_3})$. Therefore, we only calculate the decay width of this channel. And the result is
\begin{eqnarray}
&&\Gamma[X(3842)\rightarrow\chi_{_{{c2}}}(1P)\gamma]=288~\textrm{keV}.
\end{eqnarray}
We show this result and other model predictions \cite{T.Barnes2005,D.Ebert2003,B.Q.Li2009,E.J.Eichten2004,L.Cao2012,A.Parmar2010,M.A.Sultan2014} in Table \ref{II} for comparison. Where the label $NR$ represents the Non-Relativistic potential model, the Relativistic potential model is labeled $RE$; $V$ and $S$  denote the Relativistic with vector and scalar potential model; $C^3$ and $SC^3$ are Conell Coupled-channel model and Single channel potential model; $SN_{0}$ (with the zeroth-order wave functions) and $SN_{1}$ (with first-order relativistically corrected wave functions) signify Screened Nonrelativistic potential model; $CCP_{v}$ represents Coulomb plus power form of the inter-quark potential with exponent $v$.
As can be seen from Table \ref{II}, different models led to different results. Most of these results are distributed between $150\sim350$ keV. Our result, 288 keV, is in good agreement with the 296 keV of the GI model \cite{T.Barnes2005}, the 286 keV of the $C^3$ model \cite{E.J.Eichten2004}, and the 298 keV of the relativistic model \cite{M.A.Sultan2014}.

\begin{table}[!htb]
\caption{ The radiative partial decay widths (keV) of the $X_{(1^3D_{_3})}(3842)\rightarrow \chi_{_{c2}}\gamma$.}
\begin{tabular}{c|cccc}
\hline
& ~~~{\cite{T.Barnes2005}}~~ &  ~~ {\cite{D.Ebert2003}}~~  ~~ & ~~ {\cite{E.J.Eichten2004}}~~ \\
\hline
$Model$ &$NR$~~$GI$ & $NR$~~$V$~~$S$~~$RE$   & $SC^3$~~~~~~~~~~~~~~~~~~~$C^3$ \\
\hline
$M_{(1^3D_{_3})}(MeV)$ & ~~$3806$~~$3849$~~ & $3815$  & ~~$3815$~$3868$~$3972$~~~$3815$~$3868$~$3972$\\
\hline
$\Gamma(X(3842)\rightarrow \chi_{_{c2}}\gamma)$  &~~$272$~~~$296$ ~~& ~$252$~$163$~$170$~$156$~ &~~$199$~~~$329$~~~$341$~~~~$179$~~~$286$~~~$299$\\
\hline
\end{tabular}
\label{II}
\end{table}
\begin{table}[!htb]
\begin{tabular}{c|ccccc}
\hline
& ~~{\cite{B.Q.Li2009}}~~ &  ~~ {\cite{L.Cao2012}}~~ & ~~~~~~{\cite{A.Parmar2010}} ~~ & ~~ {\cite{M.A.Sultan2014}}~~&~$~ours$~ \\
\hline
$Model$ &$NR$~~$SN_{0}$~~$SN_{1}$ & ~~$NR$~~$MNR$  &~~ $CCP_{v}$ & $NR$ ~$RE$ & $BS$ \\
\hline
$M_{(1^3D_{_3})}(MeV)$ & $3799$ & $3815$~~~$3813$ & ~~$3520$~~$3653$~~$3831$ & $3805$ & $3842.7$\\
\hline
$\Gamma(X(3842)\rightarrow \chi_{_{c2}}\gamma)$~&~$272$~~~~$284$~~~~$223$~~~& ~~ $340$~~~~$302$ ~~~~&~~ $138$~~~$246$~~~$432$ ~~&~ $271$~~$298$~~ & $288$\\
\hline
\end{tabular}
\label{II2}
\end{table}

In a previous paper \cite{G.L.Wang2022}, we pointed out that in the complete relativistic method, the relativistic wave function of a state is not a pure wave. This conclusion is also valid for the charmonium \cite{W.Li2023}. For the $X(3842)$ as the $3^{--}$ state $\psi_{_3}(1^3D_{_3})$, the $D$ partial wave which survives in the nonrelativistic limit is dominant, while the $F$ and $G$ partial waves which are relativistic correction are small. While, for the $\chi_{_{c2}}(1P)$ as the $2^{++}$ state, beside the main nonrelativistic $P$ wave, it also contain a small part of relativistic $D$ and $F$ partial waves. To see these clearly, we study the contribution of different partial waves in the decay $\psi_{_3}(1^3D_{_3})\rightarrow\chi_{_{{c2}}}(1P)\gamma$. The results are presented in Table \ref{III}, where $"complete"$ means the complete or whole wave function is used, $"D~wave"$ means only the D partial wave has contribution and other partial waves are deleted.
\begin{table}[!htb]
\caption{The EM decay width (keV) of different partial waves for $X_{(1^3D_{_3})}(3842)\to\chi_{_{c2}}(1P)\gamma$.}
\begin{tabular}{c|cccc}
\hline
 \diagbox{$3^{--}$}{$2^{++}$}&$~complete~$ & $~P~wave$$(D_{_{f_5}},D_{_{f_6}})$~& $~D~wave$$(D_{_{f_1}},D_{_{f_2}},D_{_{f_7}})~$& $~F~wave$$(D_{_{f_3}},D_{_{f_4}})$ \\
 \hline
 $~complete~$ & $288$ & $215$ & $18.2$ & $0.197$\\
\hline
 $D~wave~(F_{_{5}},F_{_{6}})$ & $234$  & $232$ & $13.9$ & $0.186$ \\
\hline
 $F~wave~(F_{_{1}},F_{_{2}},F_{_{7}})$ & $13.7$ & $13.4$ & $0.180$  & $0.0187$ \\
 \hline
 $G~wave~(F_{_{3}},F_{_{4}})$ & $0.540$ & $0.245$ & $0.00373$  & $0.000358$ \\
 \hline
\end{tabular}
\label{III}
\end{table}

From the Table \ref{III}, we can see that, for the initial state, compared to $F$ and $G$ waves, the $D$ wave have the dominant contribution.  And the main contribution of the final state comes from the $P$ wave which provides the non-relativistic result, and the relativistic correction ($D$ and $F$ wave in $2^{++}$ state) contribute relatively small. Using the non-relativistic result 232 keV and the relativistic one 288 keV, we obtain the relativistic effect is $19\%$.

\section{DISCUSSION AND CONCLUSION}

In a previous paper \cite{Th.Wang2016}, we have estimated the annihilation decay (including $ggg$ and $gg\gamma$ final states) width of $X(3842)$, which is about $26.5$ keV. {From Barnes work \cite{T.Barnes2004}, we get the partial decay width $\Gamma[\psi_{_3}(^3D_{_3})\to J/\psi\pi\pi]\approx210~$keV, which is a dominant $E_1$-$E_1$ multipoles hadronic transition, and ignore other multipoles transitions which have smaller contributions compared with $E_1$-$E_1$ transition, like the $M_1$-$M_1$$\&$$E_1$-$M_2$ mode $X(3842)\to J/\psi\eta$, etc.} Then the total decay width of $X(3842)$ can be estimated as,
\begin{equation}
\Gamma[X(3842)]\approx \Gamma(DD)+\Gamma(\chi_{_{c2}}\gamma)+\Gamma(J/\psi\pi\pi)
+\Gamma(ggg)+\Gamma(gg\gamma)= 2.87~\rm{MeV}.
\end{equation}
This result is in good agreement with the experimental data $2.79^{+0.86}_{-0.86}$ MeV.

In conclusion, we study the strong and EM decays of $X(3842)$ as the $\psi_{_3(}1^{3}D_{_3})$ state by using the relativistic Bethe-Salpeter method and the $^3P_{_0}$ model. Our results are $\Gamma[X(3842)\rightarrow D^{0}\bar{D}^{0}]=1.27$ MeV, $\Gamma[X(3823)\rightarrow D^{+}D^{-}]=1.08$ MeV, $\Gamma[X(3842)\rightarrow\chi_{_{{c2}}}(1P)\gamma]=288$ keV, and the ratio $\frac{{\cal B}[X(3842)\rightarrow D^{+}D^{-}]}{{\cal B}[X(3842)\rightarrow D^{0}\bar{D}^{0}]}=0.84$. Compared with strong decay, the EM decay is not small, which is expected to be detected by experiment. In addition, we calculated the contributions of partial waves for $X(3842)\rightarrow\chi_{_{{c2}}}(1P)\gamma$, and obtained the relativistic effect $19\%$. These results may provide useful information for $X(3842)$ as the Charmonium $\psi_{_3}(^3D_{_3})$.

{\bf Acknowledgments}
This work was supported in part by the National Natural Science Foundation of China (NSFC) under the Grants Nos. 12075073, 12375085, 11865001 and 12075074, the Natural Science Foundation of Hebei province under the Grant No. A2021201009, Post-graduate's Innovation Fund Project of Hebei University under the Grant No. HBU2022BS002.

\section{APPENDIX}

\subsection{Constrained conditions of radial wave function}

For the $3^{--}$ state, we have the following relations between radial wave functions \cite{Th.Wang2016}
\begin{eqnarray}\label{3--cc}
&&f_{_1}=\frac{q^2_{_\perp}f_{_3}(\omega_{_1}+\omega_{_2})+2M^2f_{_5}\omega_{_2}}{M(m_{_1}
\omega_{_2}+m_{_2}\omega_{_1})},
~~f_{_2}=\frac{q^2_{_\perp}f_{_4}(\omega_{_1}-\omega_{_2})+2M^2f_{_6}\omega_{_2}}{M(m_{_1}
\omega_{_2}+m_{_2}\omega_{_1})},
\nonumber\\&&f_{_7}=\frac{M(\omega_{_1}-\omega_{_2})}{m_{_1}\omega_{_2}+m_{_2}
\omega_{_1}}f{_{_5}},
~~~~~~~~~~~~~~~f_{_8}=\frac{M(\omega_{_1}+\omega_{_2})}{m_{_1}\omega_{_2}+m_{_2}\omega_{_1}}{f_{_6}}.
\end{eqnarray}

For the $0^{-}$ state, we have \cite{C.S.Kim2004}
\begin{eqnarray}
&&g_{_3}=\frac{f_{_2}M_{_f}(\omega_{_{2_f}}-\omega_{_{1_f}})}{(m_{_{1_f}}\omega_{_{2_f}}+m_{_{2_f}}
\omega_{_{1_f}})},
~~~~~~g_{_4}=\frac{f_{_1}M_{_f}(\omega_{_{2_f}}+\omega_{_{1_f}})}{(m_{_{1_f}}\omega_{_{2_f}}+m_{_{2_f}}
\omega_{_{1_f}})}.
\end{eqnarray}
For the $2^{++}$ state, the relations are \cite{G.LWang2009}
\begin{eqnarray}
&&h_{_1}=\frac{(q^2_{_{f_\perp}}h_{_3}+M^2_{_f}h_{_5})(\omega_{_{1_f}}+\omega_{_{2_f}})-M^2_{_f}h_{_5}
(\omega_{_{1_f}}-\omega_{_{2_f}})}
{M_{_f}(m_{_{1_f}}\omega_{_{2_f}}+m_{_{2_f}}\omega_{_{1_f}})},
~~h_{_2}=\frac{(q^2_{_{f_\perp}}h_{_4}-M^2_{_f}h_{_6})(\omega_{_{1_f}}-\omega_{_{2_f}})}
{M_{_f}(m_{_{1_f}}\omega_{_{2_f}}+m_{_{2_f}}\omega_{_{1_f}})},
\nonumber\\&&\hspace{2cm}h_{_7}=\frac{M_{_f}(\omega_{_{1_f}}-\omega_{_{2_f}})}{m_{_{1_f}}\omega_{_{2_f}}
+m_{_{2_f}}\omega_{_{1_f}}}f_{_5},
~~~~~~h_{_8}=\frac{M_{_f}(\omega_{_{1_f}}+\omega_{_{2_f}})}{m_{_{1_f}}\omega_{_{2_f}}+m_{_{2_f}}
\omega_{_{1_f}}}f_{_6}.
\end{eqnarray}

\subsection{The positive energy wave functions}

The positive energy wave function for the $3^{--}$ state is \cite{Th.Wang2016}
\begin{eqnarray}\label{3--2}
&&\varphi^{++}_{_{3^{--}}}(q_{_\perp})=\epsilon_{_{\mu\nu\alpha}}q^{\mu}_{_\perp}q^{\nu}_{_\perp}
q^{\alpha}_{_\perp}
\bigg[F_{_1}+\frac{\slashed{P}}{M}F_{_2}
+\frac{\slashed{q}_{_\perp}}{M}F_{_3}+\frac{\slashed{P}\slashed{q}_{_\perp}}{M^2}F_{_4}\bigg]
\nonumber\\&&\hspace{2cm}+M\epsilon_{_{\mu\nu\alpha}}\gamma^{\mu}q^{\nu}_{_\perp}q^{\alpha}_{_\perp}
\bigg[F_{_5}+\frac{\slashed{P}}{M}F_{_6}+\frac{\slashed{P}\slashed{q}_{_\perp}}{M^2}F_{_7}\bigg],
\end{eqnarray}
where,
$$F_{_1}=\frac{1}{2Mm_{_1}\omega_{_1}}\bigg[q^2_{_\perp}(\omega_{_1}f_{_3}+m_{_1}f_{_4})+
M^2(\omega_{_1}f_{_5}-m_{_1}f_{_6})\bigg],$$
$$~~~~F_{_2}=\frac{M(-\omega_{_1}f_{_5}+m_{_1}f_{_6})}{m_{_1}\omega_{_1}},
~~~~F_{_3}=\frac{1}{2}(f_{_3}+\frac{m_{_1}}{\omega_{_1}}f_{_4}-\frac{M^{2}}{m_{_1}\omega_{_1}}f_{_6}),$$
$$~~~~F_{_4}=\frac{1}{2}(\frac{\omega_{_1}}{m_{_1}}f_{_3}+f_{_4}-\frac{M^{2}}{m_{_1}\omega_{_1}}f_{_5}),
~~~~F_{_5}=\frac{1}{2}(f_{_5}-\frac{\omega_{_1}}{m_{_1}}f_{_6}),$$
$$~~~~F_{_6}=\frac{1}{2}(-\frac{m_{_1}}{\omega_{_1}}f_{_5}+f_{_6}),
~~~~F_{_7}=\frac{M}{2\omega_{_1}}(-f_{_5}+\frac{\omega_{_1}}{m_{_1}}f_{_6}).$$

The positive energy wave function of the $0^{-}$ state is \cite{C.S.Kim2004}
\begin{eqnarray}\label{0-+}
&&\varphi^{++}_{_{0^{-+}}}(q_{_{f_{\perp}}})=[G_{_{f_1}}+\frac{\slashed{P}_{_{f_\perp}}}{M_{_f}}
G_{_{f_2}}+
\frac{\slashed{P}_{_f}\slashed{q}_{_{f_{\perp}}}}{M^{2}_{_f}}G_{_{f_3}}]\gamma^{5},
\end{eqnarray}
where
$$
G_{_{f_1}}=\frac{M_{_f}}{2}[\frac{\omega_{_f}}{m_{_f}}g_{_1}+g_{_2}],
~~G_{_{f_2}}=\frac{M_{_f}}{2}[g_{_1}+\frac{m_{_f}}{\omega_{_f}}g_{_2}],
~~G_{_{f_3}}=-\frac{M_{_f}}{\omega_{_f}}G_{_{f_1}}.
$$

The positive energy wave function for $2^{++}$ state $\chi_{_{c2}}$ is \cite{G.LWang2006}
\begin{eqnarray}\label{2++2}
&&\varphi^{++}_{_{2^{++}}}(q_{_{f_{\perp}}})=\epsilon_{_{f,\mu\nu}}q^{\mu}_{_{f_{\perp}}}
q^{\nu}_{_{f_{\perp}}}[H_{_{f_1}}
+\frac{\slashed{P}_{_f}}{M_{_f}}H_{_{f_2}}+\frac{\slashed{q}_{_{f_{\perp}}}}{M_{_f}}H_{_{f_3}}+
\frac{\slashed{P}_{_f}\slashed{q}_{_{f_{\perp}}}}{M^{2}_{_f}}H_{_{f_4}}]
\nonumber\\&&\hspace{2.0cm}+M_{_f}\epsilon_{_{f,\mu\nu}}\gamma^{\mu}q^{\nu}_{_{f_{\perp}}}
[H_{_{f_5}}+\frac{\slashed{P}_{_f}}{M_{_f}}H_{_{f_6}}
+\frac{\slashed{P}_{_f}\slashed{q}_{_{f_{\perp}}}}{M^{2}_{_f}}H_{_{f_7}}],
\end{eqnarray}
where
$$H_{_{f_1}}=\frac{1}{2M_{_f}m_{_f}\omega_{_f}}(\omega_{_f}q^{2}_{_{f_{\perp}}}h_{_3}
+m_{_f}q^{2}_{_{f_{\perp}}}h_{_4}
+M^{2}_{_f}\omega_{_f}h_{_5}-M^{2}_{_f}m_{_f}h_{_6}),$$
$$~~~~H_{_{f_2}}=\frac{M_{_f}}{2m_{_f}\omega_{_f}}(m_{_f}h_{_5}-\omega_{_f}h_{_6}),
~~~~H_{_{f_3}}=\frac{1}{2}(d_{_3}+\frac{m_{_f}}{\omega_{_f}}h_{_4}-\frac{M^{2}_{_f}}{m_{_f}
\omega_{_f}}h_{_6}),$$
$$~~~~H_{_{f_4}}=\frac{1}{2}(\frac{\omega_{_f}}{m_{_f}}h_{_3}+h_{_4}-\frac{M^{2}_{_f}}{m_{_f}
\omega_{_f}}h_{_5}),
~~~~H_{_{f_5}}=\frac{1}{2}(h_{_5}-\frac{\omega_{_f}}{m_{_f}}h_{_6}),$$
$$~~~~H_{_{f_6}}=\frac{1}{2}(-\frac{m_{_f}}{\omega_{_f}}h_{_5}+h_{_6}),
~~~~H_{_{f_7}}=\frac{M_{_f}}{2\omega_{_f}}(-h_{_5}+\frac{\omega_{_f}}{m_{_f}}h_{_6}).$$

\end{document}